# Noncentrosymmetric Nowotny Chimney Ladder Ferromagnet Cr$_4$Ge$_7$ with a High Curie Temperature of ~ 207 K

Zhenhai Yu[1†], Kaijuan Zhou[2†], Xiaofei Hou[1†], Xuejiao Chen[3†], Zhen Tao[2], Yunguan Ye[1], Wei Xia[1,10], Zhongyang Li[1], Jinggeng Zhao[4], Wei Wu[5], Ziyi Liu[5], Xia Wang[1,9], Na Yu[1,9], Jinguang Cheng[5], Jianlin Luo[5], Qiang Zhang[6], Vladimir Pomjakushin[7], Zhicheng Zhong[3], Soh Jian Rui[8*], Xingye Lu[2*], and Yanfeng Guo[1,10*]

[1]School of Physical Science and Technology, ShanghaiTech University, Shanghai 201210, China

[2]Center for Advanced Quantum Studies and Department of Physics, Beijing Normal University, Beijing 100875, China

[3]Key Laboratory of Magnetic Materials and Devices & Zhejiang Province Key Laboratory of Magnetic Materials and Application Technology, Ningbo Institute of Materials Technology and Engineering, Chinese Academy of Sciences, Ningbo 315201, China

[4]School of Physics, Harbin Institute of Technology, Harbin 150080, China

[5]Beijing National Laboratory for Condensed Matter Physics and Institute of Physics, Chinese Academy of Sciences, Beijing 100190, China

[6] Neutron Scattering Division, Oak Ridge National Laboratory, Oak Ridge, Tennessee 37831, USA

[7] Laboratory for Neutron Scattering and Imaging, Paul Scherrer Institut, CH-5232 Villigen, Switzerland

[8]Institute of Physics, Ecole Polytechnique Federale de Lausanne (EPFL), CH-1015 Lausanne, Switzerland

[9]Analytical Instrumentation Center, School of Physical Science and Technology, ShanghaiTech University, Shanghai 201210, China

[10]ShanghaiTech Laboratory for Topological Physics, ShanghaiTech University, Shanghai 201210, China

[†]These authors contributed equally to this work:

Zhenhai Yu, Kaijuan Zhou, Xiaofei Hou and Xuejiao Chen.

[*]Correspondence:

*jian.soh@epfl.ch;

*xingye Lu@shanghaitech.edu.cn;

*guoyf@shanghaitech.edu.cn (YFG).

Noncentrosymmetric magnets usually host intriguing magnetic interactions inherent the crystal structure with broken inversion symmetry, which can give rise to rich magnetic behaviors. We report herein the high-pressure synthesis, crystal structure, magnetizations and magnetic structure of a so-called Nowotny chimney ladder compound $Cr_4Ge_7$. Our analysis on the powder neutron diffraction data revises the crystal structure as a noncentrosymmetric space group ($P$-$4c2$, No.116). It exhibits two magnetic orders within the temperature range of 2 - 400 K. The first order at ~ 207 K associated with a small magnetic moment of ~ 0.75 $\mu_B$ is assigned to a commensurate ferromagnetic structure with a propagation vector $\mathbf{k}$ = (0, 0, 0). The weak itinerant ferromagnet nature should be caused by the complex Cr spin orders from different Wyckoff positions. The second order at ~ 18 K is assumed to arise from a competition between the Dzyaloshinskii-Moria and Heisenberg interactions. The results provide an excellent platform for study on intricate interactions between various magnetic exchanges as well as for the exploration of high temperature exotic magnetic properties which host potential applications in next-generation spintronics.

## Introduction

Symmetry breaking in solids serves as a unique venue for discovering exotic physical properties. In noncentrosymmetric superconductors such as CePt$_3$Si (*P4mm*, No. 99)[1] and K$_2$Cr$_3$As$_3$ (*P*-6*m*2, No.187), [2] the Cooper pairing is entangled with mixed spin-singlet and spin-triplet states, which result in a peculiar gap structure with point or line node. While in noncentrosymmetric magnets, the broken inversion symmetry allows an extra exchange term, i.e. the Dzyaloshinskii-Moriya (DM) interaction in the magnetic Hamiltonian. [3,4] It tends to align spins into the perpendicular orientation and therefore competes with the magnetic exchange that favours collinear spins. This will lead to different modulated chiral magnetic structures depending on the relative strength of the two interactions. Skyrmion, soliton and magnetic blue phases are some of expected chiral spin structures.[5,6] The helical magnetic skyrmions structure was observed in a small class of bulk chiral and polar magnets, such as MnSi,[7] FeGe,[8] FeCoSi,[9] Cu$_2$OSeO$_3$,[10] GaV$_4$S$_8$,[11] β-Mn-type Co-Zn-Mn[12] and VOSe$_2$O$_5$,[13] etc. This intriguing spin texture has been conceived to be capable of overcoming the limitation in conventional magnetic storage media. [14-16]

The recently discovered van der Waals (vdW) magnets offer extraordinary opportunities for the study of low-dimensional magnetism as well as for constructing various heterstructures, magnetic twistronic and spintronic devices, due to their virtue of easy exfoliation into few-layer and even monolayer.[17-22] Among various vdW magnets, the itinerant ferromagnets Fe$_n$GeTe$_2$ (n = 3, 4, 5),[23-25] Fe$_3$GaTe$_2$[26] and chromium telluride (Cr-Te)[27-34] have drawn special attentions owing to the remarkably high Curie temperature ($T_C$) which are close or even above the room temperature. A recent study unveiled that even in the centrosymmetric room-temperature Fe$_3$GaTe$_2$,[35] the introduction of iron deficiency enables spatial inversion symmetry breaking and hence a significant DM interaction, which consequently results in the appearance of room-temperature Néel-type skyrmions. The already unveiled rich properties of itinerant ferromagnets strongly suggest the necessity and urgency for the exploration of noncentrosymmetric itinerant ferromagnets, especially those with high $T_C$.

Among the various itinerant ferromagnets, the simple binary Cr-Te forms a series of self-intercalated phases including $CrTe_2$, $Cr_5Te_8$ ($Cr_{1/4}$-$CrTe_2$), $Cr_2Te_3$ ($Cr_{1/3}$-$CrTe_2$), $Cr_3Te_4$ ($Cr_{1/2}$-$CrTe_2$), and CrTe ($Cr_1$-$CrTe_2$).[33,34] A remarkable feature for these binary ferromagnets is their relatively high $T_C$ which, for example, even reaches above room temperature in $CrTe_2$ (~ 310 K),[27] $Cr_3Te_4$ (~ 320 K),[36] and CrTe (~ 340 K)[28]. As a close relative, the binary Cr-Ge system also forms abundant members including $Cr_3Ge$, $Cr_5Ge_3$, $Cr_{11}Ge_8$, CrGe, $Cr_{11}Ge_{19}$, $Cr_4Ge_7$, and $CrGe_2$ [37,38]. Unfortunately, as compared with Cr-Te, they have been less investigated. Among them, the Nowotny chimney ladder (NCL) FM $Cr_{11}Ge_{19}$ ($T_C$ ~ 88 K) crystallizes into the noncentrosymmetric $P$-$4n2$ space group (No. 118) and hosts biskyrmions spin textures associated with topological Hall effect (TIE).[39,40] It is also characterized by a rather small critical magnetic field required for the formation of biskyrmions due to its strong easy axis anisotropy associated with the characteristic helix crystal structure. These results indicate that NCL compounds may offer a unique platform to study the complex magnetic exchanges which favor intriguing topological spin textures. A very recent study on another high-pressure synthesized noncentrosymmetric ferromagnet $Cr_4Ge_7$ suggested a defect disilicide-type ($Mn_{11}Si_{19}$-type) structure and a $T_C$ of ~ 270 K.[41-43] However, the insufficient experimental data and analysis make an understanding about this intriguing phase obscure.

In this work, we report on the high-pressure synthesis, magnetizations, crystal and magnetic structures of $Cr_4Ge_7$. Based on powder neutron diffraction measurements, we clarify that $Cr_4Ge_7$ crystallizes into the $Mn_4Si_7$-type structure (space group $P$-$4c2$, No.116) rather than the $Mn_{11}Si_{19}$-type one. Furthermore, it orders with a weak itinerant ferromagnetism at ~ 207 K rather than at 270 K. We also unveil that the complex Cr orders and the competition between the DM and Heisenberg interactions play essential roles for the complex magnetic behaviors.

**Experimental**

Chromium powders (>99.9% purity) and germanium powder (~99.99% purity) were completely mixed in a molar ratio of Cr : Ge = 4 : 7 by using an agate mortar and pressed into a cylindrical ingredient. The pellets were inserted into a cylindrical hexagonal boron nitride (h-BN) capsule, and then placed into a graphite furnace. The assembly was put into an MgO octahedron and treated in a Kawai-type multi anvil high-pressure apparatus installed at ShanghaiTech University (SHTech-2000). The treatment temperature was 850 ℃ and the pressure was 5.0 GPa. After the reaction for 2 hours, the sample was quenched to room temperature without releasing the pressure. The pressure was finally gradually released at room temperature.

The crystallographic phase purity of $Cr_4Ge_7$ was initially examined on a powder X-ray diffractometer equipped with a Cu Kα radioactive source (λ = 1.5418 Å) at 298 K. Determination of the magnetic structure of polycrystalline ferromagnetic materials requires the measurements of high-flux neutron powder diffraction (NPD) at several temperatures across the magnetic transitions, because the ferromagnetic scattering is usually superposed on nuclear Bragg peaks. The high-flux time-of-flight powder diffractometer is an ideal choice. The 1st NPD measurements (Run #1) on polycrystalline $Cr_4Ge_7$ were carried out in high-resolution Time-of-Flight neutron diffractometer POWGEN, located in Oak Ridge National Laboratory. The powder sample about 2.7 g was loaded into a cylindrical vanadium container with a diameter of 6 mm. The POWGEN automatic changer was adopted as the sample environment, facilitating data collection at 7, 100, and 275 K. A neutron frame with center wavelength of 1.5 A was used to capture both nuclear and magnetic peaks across a broad $d$ spacing from 0.5 to 12.5 Å. Another independent NPD measurements (Run #2, λ = 1.886 Å) at 1 K, 50 K, 160 K, 220 K, and 290 K were performed in high-resolution powder diffractometer with thermal neutrons (HPRT) at Paul Scherrer Institut (PSI) in Switzerland. The precise structure refinement and magnetic structure analysis were performed by the Rietveld method by using GSAS with EXPGUI interface and FullProf.[44,45]

The electrical resistivity and specific heat were measured in a Quantum Design physical property measurement system (PPMS) by the standard four-electrode and relaxation methods, respectively. The Hall effect measurements were carried out using a standard Hall bar geometry in PPMS. The Seebeck and thermal conductivity, with our own designed puck and software, were measured simultaneously in the PPMS. The temperature and magnetic field dependence of magnetizations were measured in a Quantum Design magnetic property measurement system (MPMS).

The first-principles calculations were performed based on the VASP[46] with the Perdew-Burke-Ernzerhof (PBE) functional by using the plane augmented wave (PAW) method.[47] Experimental determined crystal structure parameters were input for the calculations. The Monkhorst-Pack $k$-point mesh of $11 \times 11 \times 5$ was adopted and a plane-wave cutoff energy of 500 eV was set. Spin-orbital coupling (SOC) was also included in the band structures and density of states (DOS) calculations. Different magnetic configurations were calculated and compared with the experimentally observed magnetic transitions. In order to take the correlation effec6t of Cr $d$ orbitals into account, the GGA+$U$ method was used with effective Hubbard value $U^*$ of -1 eV.[48] Here $U^*$ is defined as $U$-$J$, where $U$ is the Hubbard coulomb repulsion and $J$ is the Hund's coupling. If $J$ plays a key role, Hund's metal state may exist in the $3d$ orbital materials, which can induce mixed valence states and charge disproportionation.[49] Moreover, the $U^*$-dependent localized magnetic moments of Cr were also checked. It is found that the calculated average magnetic moments using positive value $U^*$ are always larger than the experimentally observed saturation moment. We therefore selected the negative value for calculation, which indicates that Hund's metal state may exist in $Cr_4Ge_7$, which however still need more investigations.

**Results and discussions**

The NCL phases are formed between transition metals (TM) and main group X in the formula of $TM_nX_m$, where $m$ and $n$ are integers with $1.25 \leq m/n \leq 2$. The NCL crystal structure is very abundant due to its variable chemical compositions and Wyckoff positions. Cr₄Ge₇ was ever indexed as the defect disilicides-type (Mn₁₁Si₁₉) structure,[41,43] as shown in Fig. 1(a). Fig. 1(b) shows the Mn₄Si₇-type crystal structure as a comparison, which crystallizes into the tetragonal lattice with the same NCL characteristic and has the similar lattice parameter $a$ (5.8 Å). However, the lattice parameter $c$ in Mn₁₁Si₁₉-type is about three times longer than that of Mn₄Si₇-type structure. The Mn₁₁Si₁₉-type structure exhibits a diagonal glide while the Mn₄Si₇-type one displays an axial glide. In the Mn₁₁Si₁₉-type tetragonal structure, Cr ad Ge atoms are not marked because the arrangements of Cr and Ge atoms are very complicated and they have 12 and 10 Wyckoff positions, respectively. While in the Mn₄Si₇-type tetragonal structure, as shown in Fig. 1(d), Cr atoms form a tetra-helix representing a channel with a square cross-section. Within the channel, Ge atoms form helices with a certain composition-dependent repeat distance.

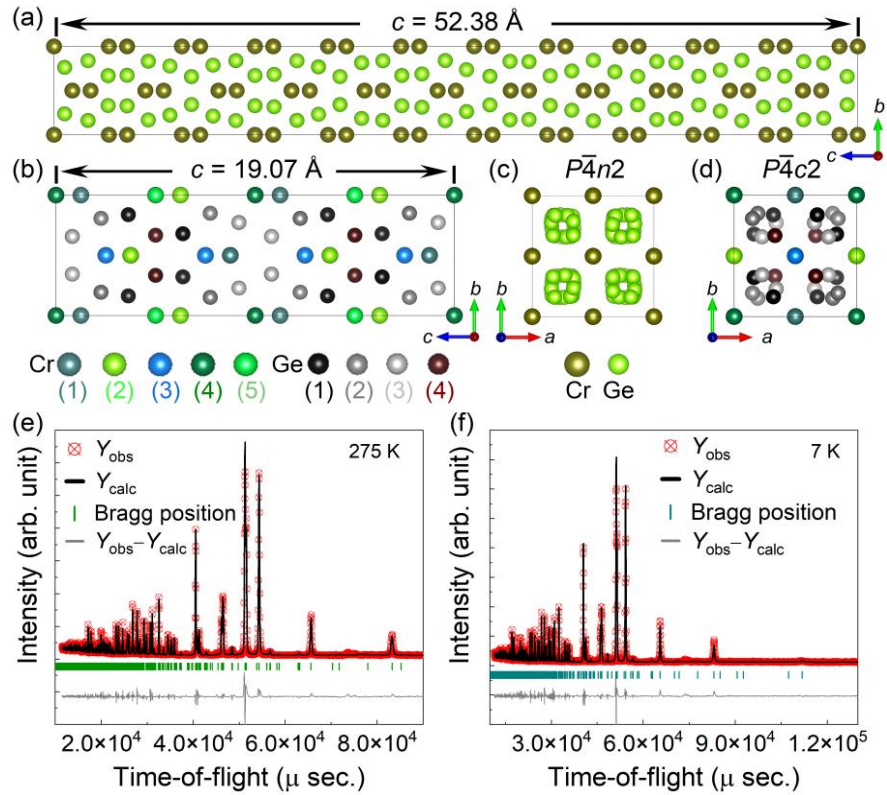

**Fig. 1**. (a)-(b) Schematic Mn₁₁Si₁₉-type and Mn₄Si₇-type NCL crystal structures viewed along the *a*-axis, respectively. (c)-(d) Arrangements of Cr and Ge atoms in the Mn$_{11}$Si$_{19}$-type and Mn$_4$Si$_7$-type NCL structures viewed along the $c$-axis, respectively. (e)-(f) Typical Rietveld refinement results of NPD for Cr$_4$Ge$_7$ at 275 and 7 K, respectively. Experimental and simulated data were symbolized with red open circle and black line, respectively. The green solid short vertical lines show positions of the allowed Bragg reflections. The difference between observed and fitted NPD patterns is shown with a gray line at the bottom of the diffraction peaks.

The analysis of powder NPD patterns of Cr$_4$Ge$_7$ collected at 275 K, as shown in Fig. 1(e), indicates a single tetragonal phase with space group $P$-4$c$2 (No.116), which corresponds to the Mn$_4$Si$_7$-type structure. This is also demonstrated by the fitting yielded lattice parameters $a = b = 5.81434(15)$ Å, $c = 19.06848(55)$ Å. The atoms in Cr$_4$Ge$_7$ show complex configurations, where Cr has five Wyckoff positions (2a, 2c, 4h, 4i, 4i) and Ge has four Wyckoff positions (4e, 8j, 8j, 8j). The detailed crystallographic information is summarized in Table S1. The crystallographic data indicate the high quality of our samples used for the following magnetic and electrical transport measurements. The refinement results at 7 K in Fig. 1(f) give the lattice parameters $a = b = 5.80371(15)$ Å, $c = 19.03735(57)$ Å, respectively.

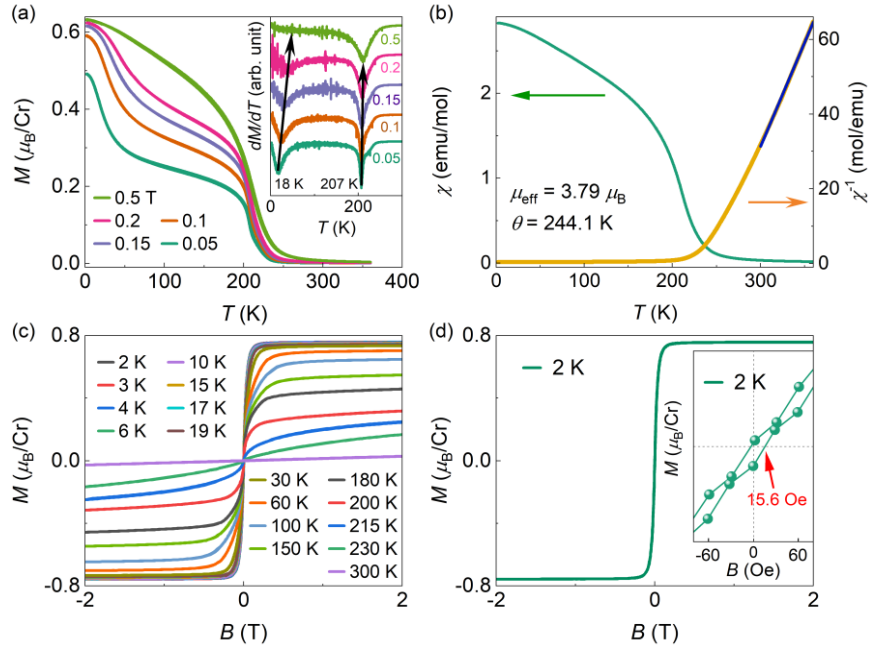

**Fig. 2.** (a) $M(T)$ at various magnetic fields. The inset shows the first order derivative of $M(T)$, where two anomalies are observed (b) $\chi(T)$ and $\chi^{-1}(T)$. (c) Magnetic field dependence of magnetic moment per Cr of $Cr_4Ge_7$ at selected temperatures. (d) The magnetic hysteresis of $Cr_4Ge_7$ at 2 K. The inset shows the enlarged view of hysteresis loop.

Fig. 2(a) shows temperature dependence of magnetization $M(T)$ under various magnetic fields. The ferromagnetic (FM) order is visible at $T_C$ of ~ 207 K as determined by the minimum of $dM/dT(T)$ curves, shown by the inset of Fig. 2(a). Besides, another anomaly is also displayed at $T_C^*$ of ~ 18 K. $T_C$ is apparently robust against the applied magnetic field, while $T_C^*$ gradually shifts to high temperatures and eventually disappears at 0.5 T, likely indicating different underlying physics of these two orders. The temperature dependent magnetic susceptibility $\chi(T)$ and its reciprocal $\chi^{-1}(T)$ are shown in Fig. 2(b). A fit of the Curie-Weiss law to $\chi^{-1}(T)$ is illustrated in Fig. 2(b), yielding the effective magnetic moment $\mu_{eff}$ of ~ 3.79 $\pm$ 0.3 $\mu_B$. Generally, in ferromagnets with localized magnetic moment, the Rhodes-Wohlfarth ratio $\mu_{eff}/\mu_{sat}$ ($\mu_{sat}$ denotes the saturation moment) is 1.0, while $\mu_{eff}/\mu_{sat}$ is larger than 1.0 in an itinerant electron ferromagnet.[50] In $Cr_4Ge_7$, our result shows that $\mu_{eff}/\mu_{sat}$ (~ 3.79 $\mu_B$/3.04 $\mu_B$) is of ~1.25, where $\mu_{sat} = 4 \times 0.76$ $\mu_B$, thus revealing a weakly itinerant character. Fig. 2(c) shows the isothermal magnetizations $M(B)$ of $Cr_4Ge_7$ at various temperatures. The saturation moment $\mu_{sat}$ decreases as the temperature increases, and the $M(B)$ curve gradually becomes a straight line above $T_C$. The data collected at 2 K is presented in Fig. 2(d), showing a $\mu_{sat}$ (~ 0.76 $\mu_B$) which is close to that of the Cr moment in $Cr_{11}Ge19$ (~ 0.82 $\mu_B$).[39] The inset of Fig. 2(d) displays an enlarged view of the hysteresis loop of $Cr_4Ge_7$, which shows a very small coercive field of ~15.6 Oe, revealing that $Cr_4Ge_7$ is a soft magnet.

To ascertain the spin structure and its evolution with the temperature across the two magnetic anomalies at ~ 18 K and ~ 207 K, analysis on the NPD data would be very helpful. Fig. 3(a) plots the NPD patterns measured at 1, 50, 160, 220 and 290 K. Since the measured data provide no indication for the presence of incommensurate magnetic peaks, we deduce that magnetic order of the Cr magnetic sublattice is commensurate with a propagation vector $\boldsymbol{k} = (0, 0, 0)$.

Symmetry analysis is performed to determine the spin configurations of the Cr magnetic sublattice, which are compatible with both $P$-$4c2$ space group and the magnetic propagation vector. We identify four irreducible representations (irreps) of spin configurations with the Cr magnetic moments oriented along the crystallographic $c$-axis. All four irreps, namely $\Gamma_1$, $\Gamma_2$, $\Gamma_3$ and $\Gamma_4$, are symmetry-distinct, with each describing a different relative orientation of the sixteen Cr moments across the magnetic sublattice.

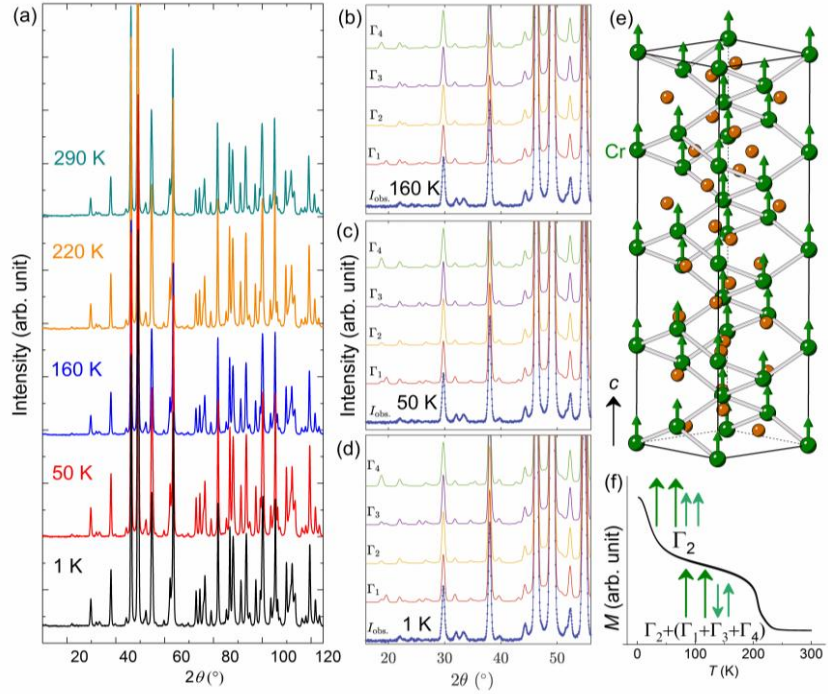

**Fig. 3.** (a) The NPD patterns (Run #2) of $Cr_4Ge_7$ at various temperatures. (b)-(d) The magnetic structures refined with the four magnetic irreps against the NPD data collected at 1, 50 and 160 K, respectively. (e) and (f) are the proposed magnetic structure and schematic magnetic phase transitions, respectively.

To simplify the analysis, we categorise the 16 Cr ions based on their Wyckoff positions, namely 4i, 4h, 2c and 2a. Table 1 tabulates the corresponding magnetic basis vectors for a given irrep, with the symbols $\pm\pm\pm\pm$ and $\pm\pm$ denoting the relative orientation of the moments for the Wyckoff positions with multiplicity 4 and 2, respectively. For instance, the $\Gamma_2$ irrep corresponds to the FM spin configuration, with all the Cr moments pointing ferromagnetically along the *c*-axis. On the other hand, irreps $\Gamma_1$, $\Gamma_3$ and $\Gamma_4$ describe antiferromagnetic (AFM) structures.

**Table 1**. Representations analysis of the Wyckoff positions of the *P*-4*c*2 space group with magnetic propagation vector ***k*** = (0, 0, 0). For instance, for the ions residing on Wyckoff position 4*i*, the basis vector ++++ indicates relative orientation of the Cr magnetic moments at the atomic positions $r_1$ = (0, ½, z), $r_2$ = (½, 0, -z), $r_3$ = (0, ½, z+½) and $r_4$ = (½, 0, -z+½), respectively.

| Irrep. / Wyckoff position | Basis vectors | | | |
|---|---|---|---|---|
| | AFM | FM | AFM | AFM |
| | $\Gamma_1$ | $\Gamma_2$ | $\Gamma_3$ | $\Gamma_4$ |
| 4i<br>(0, ½, z), (½, 0, -z), (0, ½, z+½), (½, 0, -z+½) | ++-- | ++++ | +--+ | +-+- |
| 4h<br>(½, ½, z), (½, ½, -z), (½, ½, z+½), (½, ½, -z+½) | ++-- | ++++ | +--+ | +-+- |
| 2c<br>(0, 0, 0), (0, 0, ½) | +- | ++ | | |
| 2a<br>(0, 0, ¼), (0, 0, ¾) | | ++ | +- | |

**Table 2**. The comparison of the reduced $\chi^2$ values obtained from various magnetic structures for the measured data at various temperatures. The fit was constrained to the scattering angle of $16° < 2\theta < 56°$, which accounts for all of the main magnetic peaks arising from all four irreps.

| *T* (K) | $\chi^2$ | | | | Moment ($\mu_B$) |
|---|---|---|---|---|---|
| | $\Gamma_1$ | $\Gamma_2$ | $\Gamma_3$ | $\Gamma_4$ | |
| 1 | 17.08 | **13.29** | 14.99 | 22.5 | 0.75 |
| 50 | 21.11 | **16.36** | 19.00 | 26.79 | 0.70 |
| 160 | 17.17 | **15.12** | 16.12 | 19.10 | 0.55 |

As shown in Figs. 3(b)-3(d), the four magnetic irreps against the NPD data collected at 1, 50 and 160 K are refined by using the initial structural parameters obtained at 275 K. However, due to the small Cr moments, the scattered neutron intensity is very subtle compared to the nuclear peaks. As such, for each set of data, we fix the magnitude of all the Cr moments to which obtained from the magnetization measurements presented in Fig. 2(c) and calculate the associated reduced $\chi^2$ values for the various magnetic irreps, as shown in Table 2.

We find that the $\Gamma_2$ irrep, corresponding to the FM structure, gives the best fit to the measured data at 1 K. This produces the main magnetic peaks associated with the $\Gamma_2$ irrep coinciding with the strong nuclear reflections, which are located at the Miller (*hkl*) indices such as (104), (200), (214) and (220). On the other hand, our calculations indicate that the AFM irreps ($\Gamma_1$, $\Gamma_3$ and $\Gamma_4$) should produce strong magnetic scattering intensity at reflections such as (100), (101), (103) and (114) which are not observed in the measured data. As a result, the reduced $\chi^2$ values associated with these AFM irreps are larger as compared to the FM irrep. The refinement results of the various magnetic structure models for the neutron data are also consistent with the measured temperature dependent magnetizations, indicating that Cr magnetic sublattice orders ferromagnetically below $T_C$. The proposed magnetic structure of $Cr_4Ge_7$ is shown in Fig. 3(e).

To shed light onto the evolution of the Cr spin configuration with the temperature across the transition at $T_C^* = 18$ K, we now turn to consider the NPD patterns obtained at 50 and 160 K. Interestingly, the $\Gamma_2$ irrep also provides the best fit to both sets of data, indicating that the Cr moments also order ferromagnetically in the temperature range of 18 - 207 K. This interpretation is consistent with the magnetization data. Unfortunately, due to the weak magnetic scattering intensity arising from the small Cr moment, we are unable to understand that why the magnetization with magnetic field below 0.5 T decreases on warming across 18 K. Plausible explanations include the FM configuration acquiring a small AFM component coming from the irreps $\Gamma_1$, $\Gamma_3$ and $\Gamma_4$ or spontaneous reduction of the Cr moment at ~ 18 K. To determine the evolution of the Cr magnetic order across 18 K more precisely will probably entail the use of polarized neutrons on single crystalline sample of $Cr_4Ge_7$ to benefit from the nuclear and magnetic interference, which is beyond the scope of current work. A probable spin configuration before and after the FM transition is proposed as shown in Fig. 3(f). This proposed schematic picture is based on the present measured experimental data such as characters of two "shoulders" in $M(T)$ curves and the NPD refinement.

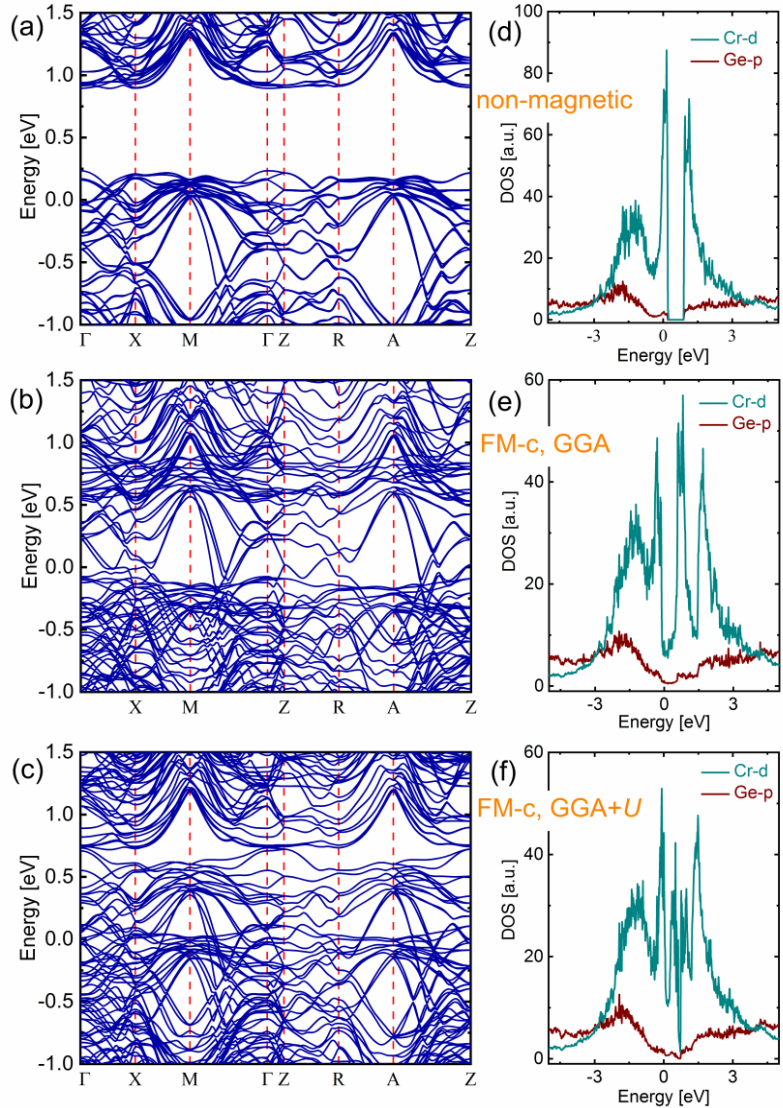

**Fig. 4.** The calculated electronic band structures of $Cr_4Ge_7$ with (a) non-magnetic state, (b) FM-c state with GGA method, and (c) FM-c state by using GGA+$U$. The bands around Fermi level are gradually renormalized due to magnetic configurations and exchange-correlation potential. The localized magnetic moment of per Cr atoms

is (b) 1.00 and (c) 0.75 $\mu_B$, respectively. The calculated DOS with SOC for (d) non-magnetic, (e) FM-c with GGA, and (f) FM-c with GGA+$U$.

The lack of an inversion center in $Cr_4Ge_7$ gives rise to DM interaction. The Heisenberg exchange tends to align the spin ordered parallel or anti-parallel, while the magnetic order is tuned to arrange vertically by DM interaction. The competition between Heisenberg exchange ($H_{EX} = -J(S_i \cdot S_j)$) and DM interaction ($H_{DM} = \mathbf{D} \cdot (S_i \times S_j)$) in noncentrosymmetric $Cr_4Ge_7$ results in the arrangement of Cr moment ordered with certain angles, which brings about complex magnetism of FM exchange with small amount of AFM component.

The density functional theory (DFT) calculations would favor a better understanding about the complex magnetic transitions in $Cr_4Ge_7$. Figs. 4(a)-4(c) present the calculated electronic band structures of non-magnetic and FM-c states respectively by using different methods. In the non-magnetic state, the Fermi level mainly goes across the top part of valence bands, in which there exists a ~ 0.67 eV gap between the valence and conduction bands, manifesting the metallic conduction in the PM state. The electronic band structure of the FM-c state calculated with GGA potential shows clear spin-splitting bands and a trend of upward shift (~ 0.5 eV) around the Fermi level, as shown in Fig. 4(b). Compared with the non-magnetic case, more dispersed bands are located at Fermi level for the FM-c spin configuration, which can improve the electrical conductivity. Furthermore, energy differences between FM-c and FM-a/FM-b are less than 1 meV for 4 × $Cr_4Ge_7$ chemical formula, which are in agreement with that the $c$-axis is the easy axis. However, the calculated average magnetic moment of Cr is 1.00 $\mu_B$, which is slightly larger than the experimental one (0.76 $\mu_B$). In order to better describe the correlated interaction of $3d$ orbitals, DFT+$U$ method is more favorable.[51,52] Compared with the GGA results, the GGA+$U$ method gives average magnetic moment per Cr of 0.75 $\mu_B$, which matches well with the experimental value. This fact indicates considerable correlation effect in $Cr_4Ge_7$. Fig. 4(c) illustrates the corresponding band structure calculated by DFT+$U$, which shows clear difference from that in Fig. 4(b) calculated by using the GGA method. As shown in Table S1, the five different Wyckoff sites Cr atoms present 0.55 (4i), 1.05 (4i), 0.84 (4h), 0.38 (2c), and 1.12 (2a) $\mu_B$, respectively. The remarkably different Cr magnetic moments on Wyckoff sites could be interpreted as the diverse magnetic interactions resulted from the crystallographic environment for the five Wyckoff positions. This means that the crystal field effect causes low and high spin states for these Cr atoms, which is close to its crystal symmetry properties. To better understand the electronic structure, we further calculate their DOS for the various cases, as shown in Figs. 4(d)-4(f). Cr-$d$ orbitals make main contribution to the magnetic properties, where peaks of DOS give itinerant ferromagnetic characteristic, matching the Stoner-type.[40] Compared with the GGA results presented in Fig. 4(e), GGA+$U$ mainly modifies the DOS distribution around the Fermi level, as shown in Fig. 4(f). These results show nice agreement with experimental observations.

Based on the above analyses, we can achieve an understanding about the FM order at 207 K. However, the anomaly at ~ 18 K remains somewhat obscure yet. Due to the iterated relaxed magnetic calculations inclined to its ground state, the unique metastable magnetic states just meet the first FM order, which is difficult to realize. In order to solve this problem, we only provide a successful converged magnetic configuration. Meantime, the calculated results with SOC or not were checked with keeping the same magnetic moment. Therefore, we neglected this SOC effect to decrease calculation cost. The converged results tell a complex AFM configuration, where the Wyckoff sites of Cr atom present AFM or FM or ferrimagnetic state. Its average magnetic moment per Cr is about 0.25 $\mu_B$. The energy difference between saturated magnetic moment and this case is 0.40 eV for 4 × $Cr_4Ge_7$ (25 meV per Cr atom). This result may be close to the real statistical magnetic configuration based on experimental measurements.

The results of electrical resistivity, magnetotransport, specific heat, and thermal transport measurements are presented in the Supplementary Information (SI), which also support the weakly itinerant ferromagnetism, the two magnetic orders and their influence on the transport properties of $Cr_4Ge_7$. The detailed information and analysis can be found therein.

**Conclusion**

In conclusion, the comprehensive studies on the noncentrosymmetric ferromagnet $Cr_4Ge_7$ revise its crystal structure as the NCL tetragonal one with a space group $P$-$4c2$. It exhibits a weak itinerant ferromagnet nature with a $T_C$ of ~ 207 K and a rather small magnetic moment of 0.75 $\mu_B$. The spins in the magnetic structure are ferromagnetically arranged along the $c$-axis, possibly mixed with certain fraction of AFM component. Besides, an anomaly is also visible at ~ 18 K. The complex magnetism of noncentrosymmetric $Cr_4Ge_7$ is originated from the competition between DM interaction and Heisenberg magnetic exchange. We would like to note that previous experimental and theoretical investigations were mainly on the noncentrosymmetric B20-type (space group: $P2_13$) ferromagnet, the study on noncentrosymmetric ferromagnets $Cr_4Ge_7$ with a different space group would yield fresh insights into the intriguing magnetic interactions and offer opportunities for the exploration of exotic magnetic properties.

**Supporting Information**

The details for magnetotransport, specific heat, and thermal transport measurements and the data analysis of the bulk Cr4Ge7 are presented in the Supplementary Information (SI).

**Data availability**

The data that support the findings of this study are available from the corresponding authors upon reasonable request.

**Competing interests**

The authors declare no competing interests.

**Acknowledgements**

The authors acknowledge the Shanghai Science and Technology Innovation Action Plan (Grant No. 21JC1402000), the National Nature Science Foundation of China (Grants No. 920651, 11934017, 12304217, 12025408) and the National Key R&D Program of China (Grants No. 2023YFA1406100, 2021YFA0718900, 2022YFA1403000, and 2021YFA1400400). Y.F.G. acknowledges the open projects from State Key Laboratory of Functional Materials for Informatics (Grant No. SKL2022), CAS and Beijing National Laboratory for Condensed Matter Physics (Grant No. 2023BNLCMPKF002). W.X. thanks the support by the open project from State Key Laboratory of Surface Physics and Department of Physics, Fudan University (Grant No. KF2022_13) and the Shanghai Sailing Program (23YF1426900). Part of this research used the Spallation Neutron Source, DOE Office of Science User Facilities operated by the Oak Ridge National Laboratory. The authors also thank the support from Analytical Instrumentation Center (#SPST-AIC10112914) and the Double First-Class Initiative Fund of ShanghaiTech University.

9141-9161.